\documentclass[10pt,letterpaper,twocolumn]{article}
\usepackage{ol}

\usepackage{amsmath,amsfonts,graphicx}

\newcommand{\beq}{\begin{equation}}
\newcommand{\eeq}{\end{equation}}
\newcommand{\beqa}{\begin{eqnarray}}
\newcommand{\eeqa}{\end{eqnarray}}

\def\pra#1{{ Phys.\ Rev. A\/} {\bf#1}}

\def\prl#1{{ Phys.\ Rev.\ Lett.} {\bf#1}}

\begin{document}

\twocolumn[

\title{Theoretical Study of Fast Light with Short sech Pulses in Coherent Gain Media}
\author{B.D. Clader and J.H. Eberly}
\affiliation{Department of Physics and Astronomy \\ University of Rochester \\ Rochester, NY 14627 USA}
%\author{J.H. Eberly}
%\affiliation{Department of Physics and Astronomy, University of Rochester \\ Rochester, NY 14627 USA}

\begin{abstract}
We investigate theoretically the phenomenon of so-called fast light in an unconventional regime, using pulses sufficiently short that relaxation effects in a gain medium can be ignored completely.  We show that previously recognized gain instabilities, including superfluorescence, can be tolerated in achieving a pulse peak advance of one full peak width.
\end{abstract}

]

\maketitle
\thispagestyle{empty}

\pagestyle{plain}
\pagenumbering{arabic}

\section{Introduction}
Studies of electromagnetically induced transparency (EIT) \cite{bib.Harris} have suggested ways that a strong laser field can dress a dielectric medium which can result in a modification of the absorptive and dispersive characteristics of the medium.  This alters the group index of the medium such that group velocity of a weak probe pulse injected in the dressed medium can be modified \cite{bib.milonniB, boyd, Grobe-etal}. Observations of slow light \cite{bib.slow} and fast light \cite{bib.fast} in such dressed media have been reported.  This modification of the medium, and subsequent change in the group velocity leads to advanced or retarded propagation of a pulse peak.  Reviews of early fast light experiments can be found in Refs. \citeonline{bib.kryukov, boyd, bib.milonniB}. 

In this paper we are concerned with fast light, which can occur if the group velocity of a pulse is greater than the vacuum speed of light, or even negative.  High dispersion allows for large change in the group index, which leads to strongly altered group velocity.  One particularly effective technique in obtaining high dispersion is to manipulate the spectral domain between two resonance lines with external {\em cw} fields.  This can create a narrow spectral window with high dispersion, through which a long probe pulse will propagate with strongly altered group velocity and little or no loss.  A steady-state high dispersion window can be created in more than one way \cite{bib.bigelow}, but the narrow bandwidth of the achieveable spectral window forces narrow-bandwidth restrictions on the probe pulse.  Noise limitations have also been discussed \cite{bib.fundquestions, Segev}.  

Coherent propagation of very short pulses is not affected by these narrow-line considerations. We will treat pulses short enough that upper level relaxation can be ignored and steady state cannot be reached.  Previous theoretical examinations of such fast-light short pulses in gain media have been reported \cite{bib.basov, bib.icsevgi}.  However because of instabilities associated with gain media, large pulse advance was not thought possible.

We have recently theoretically demonstrated \cite{Clader} that sech-shaped pulses can give a substantial pulse advance of more than 10 pulse widths even when including instabilities associated with the two-level Area theorem \cite{bib.mccall}.  This treatment did not include the quantum instability of superfluorescence, first predicted by Dicke in 1954 \cite{Dicke}, and observed in the late 1970's \cite{Vrehen, Crubellier}.  Superfluorescence theory predicts that a group of atoms will spontaneously decay collectively at a rate much greater than the single atom's decay rate.  This enhanced spontaneous emission rate causes the coherent gain medium to relax very quickly to the ground state, potentially greatly interfering with superluminal pulse propagation.  Here we describe the results of a detailed examination of fast light in a fully coherent gain medium and show that a sech-shaped input pulse still confers advantages leading to a significant pulse advance with both types of instability accounted for.

\section{Theoretical Model}
\begin{figure}[h]
\begin{center}
\leavevmode
\includegraphics[height=0.8in]{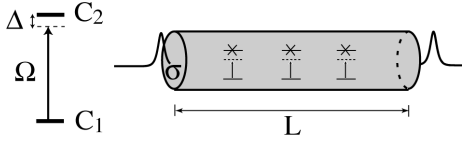}
\end{center}
\caption{\label{tla.fig} Two-level atom with level $1$ connected to level $2$ via the Rabi frequency $\Omega$ of a laser field detuned from resonance by an amount $\Delta$, and sketch of significant pulse advance in an inverted two-level gain medium of length $L$, and cross sectional area $\sigma$.}
\end{figure}

To keep our model as simple as possible, while retaining the physics of a fully coherent inverted medium, we assume the medium consists of a collection of inverted two-level atoms confined to a cell of length $L$ and cross sectional area $\sigma$ as shown in Fig. \ref{tla.fig}.  Our approach is similar to that taken by earlier workers \cite{bib.basov, bib.icsevgi}.  We allow a range of detunings $\Delta$, so the atoms are subject to inhomogeneous broadening. The Hamiltonian of a single atom in the rotating wave picture is given by
\begin{equation}
\label{eq.ham}
H = -\hbar\frac{\Omega}{2}|1\rangle\langle 2| -
\hbar\frac{\Omega^*}{2}|2\rangle\langle 1| + \hbar\Delta|2\rangle\langle 2|
\end{equation}
where $\Delta = \omega_{21} - \omega$ is the detuning of the laser below resonance, $\omega_{21}$ is the frequency difference of levels 1 and 2, $\Omega=2\mu {\cal E}/\hbar$ is the Rabi frequency associated with the $1 \to 2$ transition, $\mu$ is the dipole matrix element, ${\cal E}$ is the complex envelope function of the electric field.  The electric field has been decomposed into a slowly varying envelope function and a carrier frequency such that $E = {\cal E} e^{-i(kx - \omega t)} + \textnormal{ c.c.}$, where $\omega$ is the laser frequency, and $k=\omega/c$ is the wavenumber.  For simplicity we will restrict our attention to Rabi frequencies that are purely real for the remainder of the paper.

We assume a fully coherent atomic state of the form $|\psi (x,t)\rangle = c_1(x,t)|1\rangle + c_2(x,t)|2\rangle$ with $c_1$ and $c_2$ being the corresponding probability amplitudes.  Schr\"odinger's equation along with the Hamiltonian in Eq. \eqref{eq.ham} then gives
\begin{subequations}
\label{eq.schr}
\begin{align}
\frac{\partial c_1}{\partial t} & = i\frac{\Omega}{2} c_2 \label{schr1}
\\
\frac{\partial c_2}{\partial t} & = i\frac{\Omega}{2} c_1 - i\Delta
c_2 .\label{schr2}
\end{align}
\end{subequations}
Maxwell's wave equation written in the slowly varying envelope approximation (SVEA) yields an equation for the pulse envelope via the Rabi frequency:
\begin{align}
\label{eq.maxwell}
\left(c\frac{\partial}{\partial x} + \frac{\partial}{\partial t}\right)\Omega & = -ig \langle c_1c_2^* \rangle \\ \nonumber
& = -ig\int_{-\infty}^{\infty}d\Delta F(\Delta)c_1c_2^*
\end{align}
where $F(\Delta)=\frac{T_2^*}{\sqrt{2\pi}}e^{-(T_2^*\Delta)^2/2}$,
and $T_2^*$ is the inhomogeneous lifetime. The product $\pi gF(0)/c$ is the inhomogeneously broadened Beers absorption coefficient, and $g = N\mu^2\omega/\epsilon_0\hbar$.  Eqns. \eqref{eq.schr} and \eqref{eq.maxwell} combine to form the well known nonlinearly coupled Maxwell-Schr\"odinger (MS) system.

One can see the differences between EIT theory and this description.  We are assuming a completely coherent interaction between the laser field and the atoms by neglecting all relaxation terms in Eq. \eqref{eq.schr}. This approach is valid when the temporal width of the probe pulse and the pulse transit time is short enough to ignore relaxation.

\section{Analytic sech Pulse Solutions}
We use a mathematical approach to the coupled MS system that provides analytic expressions for fast-light pulses.  The method we use, a variant of the B\"acklund transformation technique developed by Park and Shin \cite{bib.park}, is flexible enough that we can use it to treat different possible experimental situations such as segmented and stacked media \cite{andreev}, as well as media that have a variable density of atoms.

To model a medium confined to an atomic vapor cell between positions $x_0$ and $x_1$ with length $L=x_1-x_0$ as shown in Fig. \ref{tla.fig}, we take the simplest special case for a variable density, i.e., atoms with density $N$ inside the medium, and $N=0$ outside the medium.  We will assume that the atoms in the medium are uniformly excited:  $|\psi(x,t=-\infty )\rangle = |2\rangle$, a state of perfect inversion.  The exact SVEA pulse solutions are given by the following segmented expression that describes a fully continuous pulse in time and space:
\begin{equation}
\label{eq.pulsesol}
\Omega(x,t)=
\begin{cases}
\frac{2}{\tau}\textnormal{sech}\frac{1}{\tau}\left(t-\frac{x}{c}\right)
& \text{$x < x_0$}
\\
\frac{2}{\tau}\textnormal{sech}\frac{1}{\tau}\left(t-\frac{x}{v_g}+\phi_0\right)
& \text{$x_0 \le x \le x_1$}
\\
\frac{2}{\tau}\textnormal{sech}\frac{1}{\tau}\left(t-\frac{x}{c}+\phi_1\right)
& \text{$x > x_1$},
\end{cases}
\end{equation}
where $\tau$ is the nominal temporal peak width of the pulse. The group velocity $v_g$ is given by
\begin{equation}
\label{eq.groupvel}
1- \frac{c}{v_g} = \frac{g}{2}\int_{-\infty}^{\infty} F(\Delta) \frac{d\Delta}{\Delta^2+(1/\tau)^2} \to \frac{g\tau^2}{2},
\end{equation} 
where the final limit applies when $T_2^* \gg \tau$. As the expression indicates, $v_g$ can be greater than $c$ or even negative, depending on $T_2^*$, the width $\tau$ of the pulse peak and $g$ the atom-field coupling parameter.  The expressions $\phi_0 = -\tau^2gx_0/2c$ and $\phi_1 = \tau^2g(x_1-x_0)/2c$ apply in the long $T_2^*$ limit.  The corresponding probability amplitudes for the atoms in the medium are given by 
\begin{subequations}
\label{amp_sols}
\begin{align}
c_1(x,t) & = i~\textnormal{sech}\frac{1}{\tau}\left(t-\frac{x}{v_g}+\phi_0\right)
\\
c_2(x,t) & = - \textnormal{tanh}\frac{1}{\tau}\left(t-\frac{x}{v_g}+\phi_0\right).
\end{align}
\end{subequations}

Pulse and amplitude solutions were previously obtained by Andreev \cite{andreev} for a fully coherent finite length medium in the absence of Doppler broadening. The effects of Doppler broadening have been considered for infinitely extended media \cite{Rahman-Eberly98}, and an adiabatic EIT-based example has been given by Milonni \cite{bib.milonniP}.

We define the pulse advance time as the amount by which the peak of a pulse moving through the gain medium precedes the peak of a pulse moving the same distance through vacuum. In our case we find
\begin{equation}
\label{pulsetransit}
\tau_{\textnormal{adv}} \equiv \frac{L}{c}-\frac{L}{v_g} = \frac{L}{2c}g\tau^2,
\end{equation}
where the group velocity $v_g$ in the medium is given in Eq. \eqref{eq.groupvel}. Obviously, the pulse advance time can be substantial when the group velocity is negative.

\begin{figure}[h]
\begin{center}
\leavevmode
\includegraphics[height=2.0in]{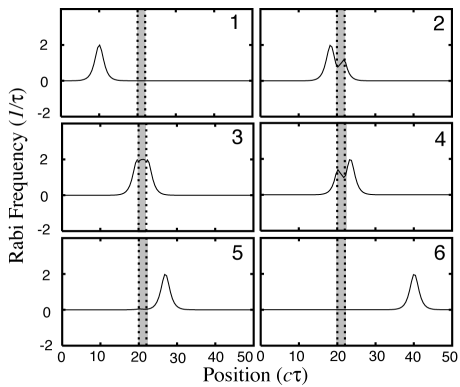}
\end{center}
\caption{\label{an_pulse.fig} Snapshots of the analytic solutions given in Eq. \eqref{eq.pulsesol}.  The horizontal axis is $x$ in units of $c\tau$.  The shaded region indicates the location of the medium which is taken to be $2c\tau$ in length.  The parameters of the pulse and medium are described in the text. For an EIT comparison, see Fig. 15 in Ref. \citeonline{bib.milonniP}.}
\end{figure}

We plot temporal snapshots of Eq. \eqref{eq.pulsesol} in Fig. \ref{an_pulse.fig}, where one can clearly see that the pulse entering the medium is advanced -- in the third frame an outgoing peak leaves the medium even before the input peak enters it.  In terms of initial temporal peak widths the peak advance is $\tau_{\textnormal{adv}} \approx 2.66 \tau$ for a sample chosen to be Rb vapor, using the D$_2$ line with these parameters: $T_2^* = 0.733$ ns, $g = 266$ ns$^{-2}$ and $\tau = 0.1$ ns, all of which appear to be within experimental reach.  These parameters give an inverse Beer's length of $\alpha = \sqrt{\pi/2}gT_2^*/c \approx 8.15$ cm$^{-1}$ and a negative group velocity in the medium of $v_g/c \approx -3.27$.  The medium was taken approximately 50 gain lengths deep, which corresponds to about 6 cm.  Clearly Eq. \eqref{pulsetransit} predicts that the pulse advance time can be made arbitrarily large by simply increasing the length of the medium.

\section{Edge-Effect Instabilities}

After interacting with the medium, the analytic pulse solutions given in Eq. \eqref{eq.pulsesol} return the atoms to their excited state, which is easily seen in Eq. \eqref{amp_sols} where $|c_2(x,t \to \pm \infty)| = 1$. This is a fully coherent but unstable situation.  We cannot realistically assume infinitely long pulse wings (as in the analytic sech functions). Since the studies of Sommerfeld and Brillouin \cite{Oughston-Sherman}, the effects of finite cutoffs at the leading and also trailing edges of the pulse have been recognized as important to take into account. Instabilities will always arise in gain media from such effects \cite{Lamb, Gabitov-etal2}.  These instabilities can be understood via  the two-level Area theorem \cite{bib.mccall}.  If one defines the Area of the pulse to be
\begin{equation}
\label{pulse_area}
\theta(x) = \int_{-\infty}^{\infty} \Omega(x,t) dt,
\end{equation}
then the well known Area theorem for an inhomogeneously broadened gain medium (i.e. $T_2^* \ll \tau$) is
\begin{equation}
\label{area_theorem}
\frac{\partial \theta(x)}{\partial x} = \frac{\alpha}{2} \sin\theta(x),
\end{equation}
where $\alpha = \pi g F(0)/c$ is the inverse Beers length.  From Eq. \eqref{area_theorem} we can clearly see that pulses of Area $2\pi$ are unstable.  The analytic pulse solutions have Area = $2\pi$, so only the perfect infinitely long analytic solutions are stable.  Any slight variation, such as sharply cut off leading and trailing edges, will cause the pulse shape to change as the Area  tends towards the stable values of $\pi$ or $3\pi$.

\begin{figure}[h]
\begin{center}
\leavevmode
\includegraphics[height=0.8in]{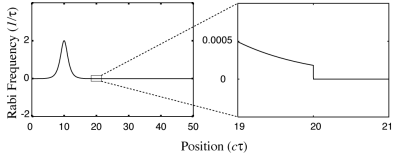}
\end{center}
\caption{\label{pulseEdge}The left frame shows the sech pulse used in numerical simulations.  The right frame zooms in on the front edge of the pulse showing a sharp front edge 10 units from the pulse peak.  The horizontal and vertical axes are the same as those in the analytic and numerical solution figures.}
\end{figure}

We tested the sensitivity of the semiclassical equations to the cutoff instability by generating additional numerical solutions to the same set of equations, with the same medium parameters as in the previous section.  We used a short enough pulse width to justify ignoring significant medium relaxation, but we gave the pulse weak but sharp leading and trailing cutoffs as shown in Fig. \ref{pulseEdge}.  In the previous section the analytic formulas predict a pulse advance time of about 2.5 peak widths for the parameters chosen, so we inserted cutoffs 10 pulse widths from the pulse peak.

\begin{figure}[h]
\begin{center}
\leavevmode
\includegraphics[height=2.0in]{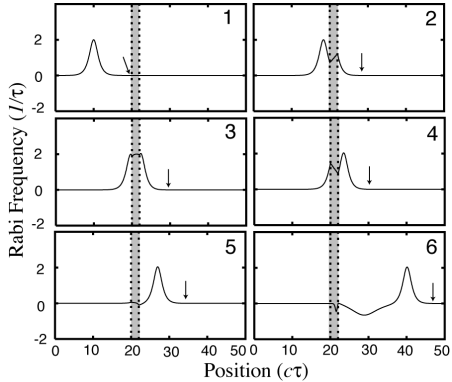}
\end{center}
\caption{\label{num_pulse.fig} Snapshots during evolution of the numerical pulse solution to Eqns. \eqref{eq.schr} and \eqref{eq.maxwell} with the same parameters as for Fig. \ref{an_pulse.fig}, but for a pulse initially entirely outside the medium with abrupt edges at $\pm$ 10 units from the peak.  The pulse peak advances $\sim 2.5$ units forward, which is still well behind the front edge of the pulse, which is marked by the arrows.}
\end{figure}

Snapshots of the numerical solution are shown in Fig. \ref{num_pulse.fig}, and in frame 3 one sees an advanced peak leaving the medium before the input peak enters, along with a back-moving pulse (negative group velocity) in the medium, just as in the analytic solution. The output pulse is relatively cleanly separated from background, and is essentially identical to that for the analytic solution, in both its extent of advance and in its Area which is $2\pi$.  The negative pulse envelope, a kind of pulse ``ringing" that trails the advanced pulse, is radiation arising from the coherence stimulated in the medium by the leading and trailing edges of the pulse.

\begin{figure}[h]
\begin{center}
\leavevmode
\includegraphics[height=1.5in]{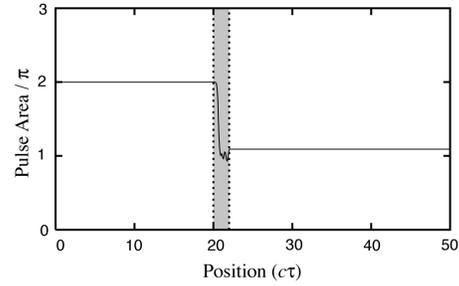}
\end{center}
\caption{\label{num_area}Time integrated area of the pulse from the numerical solution shown in Fig. \ref{num_pulse.fig}.  The horizontal axis is position in units of $c\tau$ and the vertical axis is the area of the pulse, as given in Eq. \eqref{pulse_area} in units of $\pi$.}
\end{figure}

The Area of the pulse calculated from the numerical solution is plotted in Fig. \ref{num_area}.  As the pulse enters the medium, its Area quickly changes from the unstable $2\pi$ to the stable Area $\pi$.  However the Area does not settle to $\pi$ but rather oscillates slightly about that point.  This is due to the fact that inhomogenous broadening does not dominate (cf. rephasing described by Burnham and Chiao \cite{burnham-chiao}).  This Area change can also be seen in Fig. \ref{num_pulse.fig} where the analytic solution with Area $2\pi$ is followed by the pulse ringing with Area oscillating about $-\pi$.

Thus, in spite of a cutoff instability, we have shown that the analytic pulse formulas in Eq. \eqref{eq.pulsesol} and the analytic pulse advance time in Eq. \eqref{pulsetransit} are still valid as long as the peak advance does not get ahead of the sharp front edge.  After the pulse leaves the medium the instabilities appear as pulse ringing trailing the advanced pulse.

\section{Effect of Spontaneous Instabilities}

In addition to the cutoff instabilities mentioned in the previous section, purely quantum processes provide a different source for instability. Spontaneous emission is the most fundamental, but single-atom spontaneous lifetimes are rather long, say 10's to 100's of nanoseconds, which we have so far overlooked. However, a more important quantum process is superfluorescence (SF), i.e., collective spontaneous emission, for which the excited state lifetime is dramatically reduced, perhaps into the regime where it can't be ignored. Specifically, a superfluorescent pulse will emerge after a short delay time given roughly by $\tau_{\textnormal{D}} \sim \pi\tau_0/(4N\lambda^2L)$ for a sample with Fresnel number $F \sim 1$. Here $N$ is the density of atoms in the sample, $\tau_0$ is the single atom radiative decay time due to spontaneous emission, $\lambda$ is the wavelength of the emitted radiation, and $L$ is the length of the sample \cite{Bonifacio, Gabitov-etal}.  This expression decreases with increasing length $L$, as should be expected, whereas the pulse advance time given in (\ref{pulsetransit}) increases. Since successful observation of peak advance can be realized only if the SF pulse develops after pulse advance has occurred, this presents a conflict that must be accommodated.

\begin{figure}[h]
\begin{center}
\leavevmode
\includegraphics{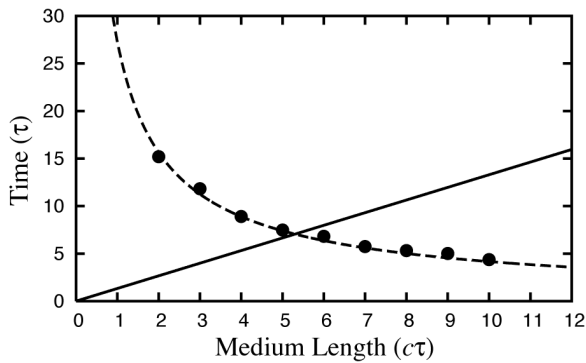}
\end{center}
\caption{\label{dtime.fig} Plot of the delay time, $\langle\tau_D\rangle$ of the superfluorescent pulse given in Eq. \eqref{SFDelay} and the total pulse transit time given in Eq. \eqref{pulsetransit}.  The horizontal axis is the length of the medium in units of $c\tau$ and the vertical axis is time in units of $\tau$.  The dashed line is $\langle\tau_D\rangle$ and the solid line is $\tau_{\textnormal{adv}}$.  Each data point is an average delay time from 20 numerical solutions, showing excellent agreement with the theoretical delay time in Eq. \eqref{SFDelay}.  Pulse advance is predicted for a medium of up to $5c\tau$, giving pulse advances of up to 6.5 $\tau$. }
\end{figure}

To examine this conflict we have used the results of Polder \textit{et al.} \cite{Polder} who calculated the average delay time (average time for an SF pulse to emerge) to be given by
\begin{equation}
\label{SFDelay}
\langle\tau_D\rangle = \frac{\pi\tau_0 V}{4N_a\lambda^2L}\big(\ln(2\pi N_a)\big)^2=\frac{3c}{4gL}\left[\ln\left(\frac{L}{L_0}\right)\right]^2, 
\end{equation}
where $L_0=(2\pi N\lambda)^{-1/2}$.   We plot both expressions (\ref{pulsetransit}) and (\ref{SFDelay}) in Fig. \ref{dtime.fig} as functions of medium length, in units of peak width $c\tau$. The result is positive, in the sense that there is a range of medium lengths up to about $L \sim 5c\tau$ for which the SF time is longer than the advance time. For these times we can expect to see pulse advances up to about 6.5 peak widths.  To ensure that the SF pulse delay is substantially longer than the pulse advance time, shorter lengths may be preferred, as we have chosen in our examples.

\begin{figure}[h]
\begin{center}
\leavevmode
\includegraphics[height=2.0in]{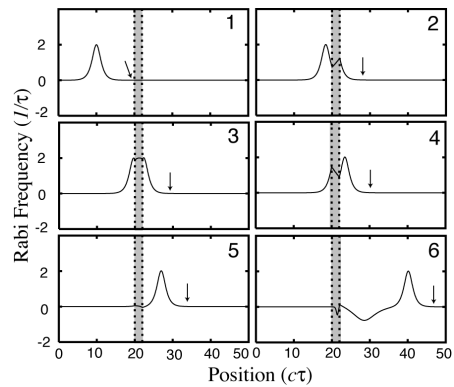}
\end{center}
\caption{\label{PAdv_SF.fig} Snapshots during evolution of the numerical pulse solution to Eqns. \eqref{eq.schr} and \eqref{eq.maxwell} with the same parameters as for Fig. \ref{an_pulse.fig}, but with initial probability amplitudes as given in Eq. \eqref{sfinitial}. The pulse is initially entirely outside the medium with sharp edges at $\pm$ 10 units from the peak, with the front edge indicated by the arrows.  The following peak advance is ~2.5 units forward, in agreement with Eq. \eqref{pulsetransit}.  This advanced pulse  is also ahead of the superfluorescent pulse shown in Fig. \ref{SF.fig}.}
\end{figure}

Spontaneous processes can be included in our semiclassical treatment by adding an incoherent stochastic trigger to the MS equations. We have done this via zero-average fluctuating dipole moments as a surrogate for the vacuum fluctuations that are the fundamental basis for spontaneous emission.  This modification is dealt with in detail in the Appendix. We then tested for pulse advance in the presence of both types of instability by solving numerically the same MS system, equations \eqref{eq.schr} and \eqref{eq.maxwell}, in this instance with edges cut off, and also with random initial dipole moments (see Eqn. \eqref{sfinitial}). We have retained the same finite length $2\pi$ sech input pulse described in Sec. IV.  We used a medium $2c\tau$ in length (6 cm for a 100 ps pulse) just as in the previous examples. Actually all of the cases illustrated in this paper employ parameters chosen in advance with guidance from the Appendix so as to satisfy the limitation that SF places on the length of the medium.   The solution is plotted in Fig. \ref{PAdv_SF.fig}, and we see almost no difference between that and Fig. \ref{num_pulse.fig} where we assumed a perfect initial inversion. The pulse advance in this example is about $~2.5$ pulse widths in agreement with Eq. \eqref{pulsetransit}, and the advanced pulse exits the medium before a subsequent ``ringing" pulse.  The almost identical features of the solutions in Figs. \ref{num_pulse.fig} and \ref{PAdv_SF.fig} indicate that SF plays practically no role under the conditions chosen.

\section{Summary}

We have examined theoretically the propagation of sech pulses in an idealized gain medium treated as inverted two-level atoms.  The pulses are taken short enough to ignore collisional and radiative lifetimes for the atoms. Our motivation has been to test the feasibility of achieving substantial peak advance (so-called ``fast light") in transmission through the medium, and here substantial means one full peak width or more.  Fully coherent propagation of a resonant pulse in weakly Doppler-broadened Rb vapor was used throughout as the basis for our modelling. This is an example for which resonant input pulses are readily available experimentally with durations much shorter than the 26 ns spontaneous lifetime associated with the D2 transition.

We exhibited an analytic solution to the nonlinearly coupled semiclassical Maxwell-Schr\"odinger equations in terms of a pulse with $2\pi$ Area, and showed in Sec. III that this solution is a ``fast light" example with a peak advance that can theoretically be made arbitrarily large by adjusting the length of the gain medium.  However we then included pulse cutoffs to introduce edges (which can travel no faster than the vacuum velocity). Nevertheless, a numerical solution carried out under the conditions employed for the analytic sech pulse showed that when these deviations are introduced, the fast light result is still obtained, and with nearly the same peak advance. A train of irregular pulses was induced by the edges, but they followed and did not interfere with the advanced peak. Finally, with the aid of modelling described in the Appendix, we took into account purely quantum effects. We used stochastic dipole fluctuations to induce collective spontaneous emission throughout the medium, i.e., superfluorescence, and found that it in our chosen regime it does not alter the peak advance in a significant way.  

In summary, we have extended the results obtained in previous thereortical examinations of fast light in fully coherent gain media. The combination of results presented here suggests that previous conclusions were unnecessarily pessimistic. Since the effect of superfluorescence is proportional to the number of participating atoms, it is desirable to restrict the length of the medium, and in the Rb vapor example used, we could show that even in the presence of SF a peak advance of over one peak width could be arranged for transmission of a 100 ps sech pulse in a 6 cm (50 absorption depths) medium.  Thus we predict that a very short input pulse can provide access to fast light using pulse bandwidths several orders of magnitude greater than has been utilized to date, or projected to be possible via EIT.

\section*{Acknowledgments}
We wish to thank Q-Han Park, P.W. Milonni, and I.R. Gabitov for useful discussions.  B.D. Clader acknowledges receipt of a Frank Horton Fellowship from the Laboratory for Laser Energetics, University of Rochester. Research supported by NSF Grants PHY 0456952 and 0601804.

\appendix

\section{Maxwell-Schr\"odinger Model of Superfluorescence}
To confirm that our conclusions in Sec. V satisfactorily incorporate quantum fluctuations, we have compared results from our modeling with established theoretical studies of superfluorescence (SF) that have been backed up by experimental tests. The MS system that we are using assumes a classical field.  However SF can be extended to the MS system by appropriately modeling the quantum fluctuations with zero-average fluctuating dipole moments in the inverted medium \cite{Polder, MacGillivray, Haake}.  Instead of perfect inversion, we assume the medium is initially prepared as
\begin{align}
\label{sfinitial}
c_1(x_i,t=-\infty) & = \sin\left(\frac{\theta_0(x_i)}{2}\right)e^{i\phi(x_i)} \nonumber
\\
c_2(x_i,t=-\infty) & = \cos\left(\frac{\theta_0(x_i)}{2}\right),
\end{align}
where $\theta_0(x_i)$ is a randomly chosen very small initial tipping angle of the Bloch vector, and $\phi(x_i)$ is a random phase, for each point $x_i$.  For simplicity we assign $\phi(x_i) = 0 \textnormal{ or } \pi$ randomly to give a zero average complex dipole moment, $\mu\langle c_1c_2^*\rangle = 0$, when averaged over the length of the medium.  Arbitrary phases in our numerical solutions do not change the results significantly, in agreement with previous studies \cite{MacGillivray}.

The modeling procedure assigns to the tipping angle $\theta_0(x_i)$ a very small randomly chosen number with gaussian statistics at each point in the medium $x_i$.  The average value of the tipping angle is given by $\langle\theta_0\rangle = 2/\sqrt{N_a}$ with variance $\langle\theta_0^2\rangle - \langle\theta_0\rangle^2 = 1/N_a$, where $N_a$ is the number of atoms and the average is over the length of the medium.  We assume an initial density of atoms of $N = 8 \times 10^{10} \textnormal{ cm}^{-3}$ confined to a cylindrical atomic vapor cell with Fresnel number $F = \sigma/\lambda L = 1$.  Thus the number of atoms is $N_a = N\sigma L =  N\lambda L^2 \sim 10^8$, for a 6 cm vapor cell. Since SF is a purely quantum mechanical effect, theoretical studies have derived these statistics for the initial tipping angle to ensure that the semi-classical MS system gives the same superfluorescent delay time (defined below) as purely quantum mechanical derivations \cite{MacGillivray, Polder, Haake}. Subsequent experimental measurements confirmed these theoretical results \cite{Vrehen, Crubellier}.

Polder \textit{et al.} \cite{Polder} define the superfluorescent delay time, $\tau_D$, to be the time for the tipping angle in Eq. \eqref{sfinitial} at the output face of the medium, $x=x_1$, to grow to become equal to 1.  The SF delay time has alternative definitions, and has been derived in a variety of other ways \cite{Gabitov-etal, MacGillivray}, and the statistics have been studied \cite{Haake}.  The delay time specifically refers to the time it takes for the SF pulse to be emitted.  However it can also be regarded as a decay time since, as the atoms emit the SF pulse, they subsequently decay to the ground state.  

\begin{figure}[h]
\begin{center}
\leavevmode
\includegraphics[height=2.0in]{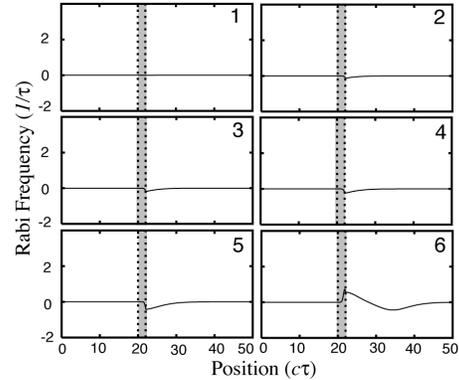}
\end{center}
\caption{\label{SF.fig} Snapshots during evolution of the numerical pulse solution to Eqns. \eqref{eq.schr} and \eqref{eq.maxwell} but with no input pulse, and with initial probability amplitudes as given in Eq. \eqref{sfinitial}. A superfluorescent pulse emerges from the medium with delay time $\langle \tau_D\rangle$.  The time of each snapshot is the same as in all other figures..}
\end{figure}

We use the expression (\ref{SFDelay}) of Polder \textit{et al.} for average delay time. To confirm that our numerical solutions conform to this model for superfluorescence, we have averaged the delay times observed in 20 ``experiments", i.e., for a given length $L$ we made 20 different numerical solutions of Eqns. \eqref{eq.schr} and \eqref{eq.maxwell} using as an initial condition Eq. \eqref{sfinitial}, but without any input pulse. An example of one of the solutions with a delay time of about $\langle\tau_D\rangle$ is shown in Fig. \ref{SF.fig}. This procedure was repeated for 9 different lengths $L$, and the 9 averages are plotted as the data points shown in Fig. \ref{dtime.fig}. One sees excellent agreement between the average delay time from our numerical SF model and the expression (\ref{SFDelay}) for the average delay time.

\begin{thebibliography}{99}

\bibitem{bib.Harris} S.E. Harris, ``Electromagetically Induced Transparency," Phys. Today \textbf{50}, 36-42 (1997).

\bibitem{bib.milonniB} P.W. Milonni, \textit{Fast light, slow light
and left-handed light} (Institute of Physics Publishing, Bristol,
2005).

\bibitem{boyd} R.W. Boyd and D.J. Gauthier, ``Slow and Fast Light," Prog. Opt. \textbf{43}, 497-530 (2002).

\bibitem{Grobe-etal} R. Grobe, F.T. Hioe and J.H. Eberly, ``Formation of Shape-Preserving Pulses in a Nonlinear Adiabatically Integrable System," \prl{73}, 3183-3186 (1994).

\bibitem{bib.slow} A. Kasapi, M. Jain, G.Y. Yin, and S.E. Harris, ``Electromagnetically Induced Transparency: Propagation Dynamics," \prl {74}, 2447-2450 (1995).

\bibitem{bib.fast} L.J. Wang, A. Kuzmich, and A. Dogariu, ``Gain-assisted superluminal light propagation, "Nature
(London) \textbf{406}, 277-279 (2000).

\bibitem{bib.kryukov} P.G. Kryukov and V.S. Letokhov, ``Propagation of a light pulse in a resonantly amplifying (absorbing) medium," Sov. Phys. Usp. \textbf{12}, 641-672 (1970).

\bibitem{bib.bigelow} M.S. Bigelow, N.N. Lepeshkin, and R.W. Boyd, ``Superluminal and Slow Light Propagation in a Room-Temperature Solid," Science \textbf{301}, 200202 (2003).

\bibitem{bib.fundquestions} M.D. Stenner, D.J. Gauthier, and M.A. Neifeld, ``The speed of information in a `fast-light' optical medium," Nature \textbf{425}, 695-698 (2003).

\bibitem{Segev} B. Segev, P.W. Milonni, J.F. Babb, and R.Y. Chiao, ``Quantum Noise and Superluminal Propagation," \pra{62}, 022114 (2000).

\bibitem{bib.basov} N.G. Basov, et al., ``Nonlinear amplification of light pulses," Sov. Phys. JETP \textbf{23}, 16-22 (1966).

\bibitem{bib.icsevgi} A. Icsevgi and W.E. Lamb, Jr., ``Propagation of Light Pulses in a Laser Amplifier," Phys. Rev.
\textbf{185}, 517-545 (1969).

\bibitem{Clader} B.D. Clader, Q-Han Park, and J.H. Eberly, ``Fast Light in Fully Coherent Gain Media," Opt. Lett. \textbf{31}, 2921-2923 (2006).

\bibitem{bib.mccall} S.L. McCall and E.L. Hahn, ``Self-Induced Transparency," Phys. Rev. \textbf{183}, 457-485 (1969).

\bibitem{Dicke} R.H. Dicke, ``Coherence in Spontaneous Radiation Processes," Phys. Rev. \textbf{93}, 99-110 (1954).

\bibitem{Vrehen} Q.H.F. Vrehen and M.F.H. Schuurmans, ``Direct Measurement of the Effective Initial Tipping Angle in Superfluorescence," \prl{42}, 224-227 (1979).

\bibitem{Crubellier} A. Crubellier, S. Liberman, and P. Pillet, ``Superradiance Fluctuations in a $j=\frac{1}{2} \to j\prime = \frac{1}{2}$ Atomic System," J. Phys. B \textbf{17}, 2771-2780 (1984).

\bibitem{bib.park} Q-Han Park and H.J. Shin, ``Matched pulse propagation in a three-level system," Phys. Rev. A \textbf{57}, 4643-4653 (1992).

\bibitem{andreev} A.V. Andreev, ``Optical superradiance: new ideas and new experiments," Sov. Phys. Usp. \textbf{33}, 997-1020 (1990).

\bibitem{Rahman-Eberly98} A. Rahman and J.H. Eberly, ``Theory of shape-preserving short pulses in inhomogeneously broadened three-level media," \pra{58}, R805-R808 (1998).

\bibitem{bib.milonniP} P.W. Milonni, ``Controlling the speed of light pulses," J. Phys. B \textbf{35}, R31-R56 (2002).

\bibitem{Oughston-Sherman} For a discussion and extension of the pioneering works of Sommerfeld and Brillouin, see, for example, {\em Electromagnetic Pulse Propagation in Causal Dielectrics}, by K.E. Oughston and G.C. Sherman (Springer-Verlag, Berlin-Heidelberg, 1994) Corrected Edition (1997).

\bibitem{Lamb} G.L. Lamb, Jr., ``Amplification of coherent optical pulses," \pra{12}, 2052-2059 (1975).

\bibitem{Gabitov-etal2} I.R. Gabitov and S.V. Manakov, ``Propagation of Ultrashort Optical Pulses in Degenerate Laser Amplifiers," \prl{50}, 495-498 (1983).

\bibitem{burnham-chiao} D.C. Burnham and R.Y. Chiao, ``Coherent Resonance Fluorescence Excited by Short Light Pulses," Phys. Rev. \textbf{188}, 667-675 (1969).

\bibitem{Gabitov-etal} I.R. Gabitov, V.E. Zakharov, and A.V. Mikhailov, ``Maxwell-Bloch equation and the inverse scattering method," Theor. Math. Phys. \textbf{63}, 328-343 (1985).

\bibitem{Bonifacio} R. Bonifacio and L.A. Lugiato, ``Cooperative Radiation Processes in two-level Systems: Superfluorescence," \pra{11}, 1507-1521 (1975).

\bibitem{Polder} D. Polder, M.F.H. Schuurmans, and Q.H.F. Vrehen, ``Superfluorescence: Quantum-mechanical derivation of Maxwell-Bloch description with fluctuating field source," \pra{19}, 1192-1203 (1979).

\bibitem{MacGillivray} J.C. MacGillivray and M.S. Feld, ``Theory of superradiance in an extended, optically thick medium," \pra{14}, 1169-1189 (1976).

\bibitem{Haake} F. Haake, J.W. Haus, H. King, and G. Schr\"oder, ``Delay-time statistics of superfluorescent pulses," \pra{23}, 1322-1333 (1981).

\end{thebibliography}
\end{document}